\documentclass[twocolumn,superscriptaddress,aps,prd,nolongbibliography,floatfix]{revtex4-2}

\usepackage{amsmath}
\usepackage{amssymb}
\usepackage{graphicx}
\usepackage{dcolumn}
\usepackage{xcolor}
\usepackage{bm}
\usepackage{hyperref}

\usepackage[normalem]{ulem}
\usepackage{epstopdf}
\pdfoutput=1

\newcommand{\istd}{\mathrm{i}}
\newcommand{\dstd}{\mathrm{d}}

\newcommand{\ea}{{\it et al.}}

\newcommand{\eff}{\mbox{E$_{\text{field}}$}}

\begin{document}


\title{Electric field control of magnons in magnetic thin films:\\
ab initio predictions for 2D metallic heterostructures}

\author{Alberto Marmodoro}
 \email{marmodoro@fzu.cz}
 \affiliation{FZU – Institute of Physics of the Czech Academy of Sciences, Cukrovarnick\'{a} 10, CZ-162~53~Prague, Czech Republic}
\author{Sergiy Mankovsky}
\author{Hubert Ebert}
 \affiliation{Department of Chemistry, Ludwig-Maximilians-University Munich, Butenandtstrasse 11, D-81377 Munich, Germany}
\author{Jan Minár}
\affiliation{
New Technologies Research Centre, University of West Bohemia, 
CZ-301~00~Pilsen, Czech Republic
}
\author{Ond\v{r}ej \v{S}ipr}
 \affiliation{FZU – Institute of Physics of the Czech Academy of Sciences, Cukrovarnick\'{a} 10, CZ-162~53~Prague, Czech Republic}
\affiliation{
New Technologies Research Centre, University of West Bohemia, 
CZ-301~00~Pilsen, Czech Republic
}

\date{\today}

\begin{abstract}
We explore possibilities for control of magnons in two-dimensional
heterostructures by an external electric field acting across a
dielectric barrier.  By performing ab-initio calculations for a Fe
monolayer and a Fe bilayer, both suspended in vacuum and deposited on
Cu(001), we demonstrate that external electric field can significantly
modify magnon lifetimes and that these changes can be related to
field-induced changes in the layer-resolved Bloch spectral functions.
For systems with more magnon dispersion branches, the gap between
high- and low-energy eigenmodes varies with the external field.  These
effects are strongly influenced by the substrate.  Considerable
variability in how the magnon spectra are sensitive to the external
electric field can be expected, depending on the substrate and on the
thickness of the magnetic layer.

\end{abstract}

\pacs{Valid PACS appear here}
\maketitle


\section{Introduction}
Magnonics, i.e. the generation, control and detection 
of collective spin excitations (or magnons)
is been considered for possible
information storage and processing applications,
due to promise for higher data density 
and its more energy-efficient elaboration
\cite{Demokritov2013,Chumak2015,Tannous2015,Zakeri2018,Mahmoud2020,Xu2020}.
This area is rapidly advancing, 
from first proposals of memory devices, 
to more recent examples concerning the implementation of logical
operations \cite{Kostylev2005,Guo2018,Wang2020}.

Various groups have studied 
how an external electric field 
can be used to modify features of the magnon spectra 
and to potentially realize these functionalities. 
An early example has been the measurement of proportionality between
magnetic resonance shifts and an applied electric field in lithium ferrite \cite{Rado1979}.
This observation has been explained as a consequence of a voltage-controlled magneto-crystalline anisotropy (VCMA) variation, and deemed small for practical applications \cite{Liu2013}.  
Subsequently, multiferroic materials have been found to offer a stronger response
in their magnon spectrum
through the coupling between their intrinsic electric polarization 
and the externally applied perturbation \cite{Rovillain2010,Risinggard2016}.
More recently, Liu {\it et al.} have discussed yet a different theoretical mechanism
not restricted to this class of materials
and capable to produce effective Dzyaloshinskii-Moriya interactions
(DMI) proportional to the field \cite{Liu2011b}.
This has prompted to examine implications for magnon spectra
\cite{Zhang2014a,Krivoruchko2017a,Krivoruchko2018,Rana2019,Savchenko2019b,Krivoruchko2020},
most frequently adopting as reference material 
the ferrimagnetic insulator yttrium iron garnet (YIG).

In this work we are interested in the possible
control of magnons by an applied electric field acting, 
across a dielectric barrier, on a two-dimensional (2D) 
heterostructure.
We deal with the idealized layout 
of magnetic/non-magnetic layers of simple
transition metals, e.g. Fe and Cu. 
Similarly to the case of YIG, absence of electric current 
due to the insulating barrier
precludes energy dissipation into Joule heating (Ohmic losses).
The gating E$_{\textrm{field}}$ acts 
by controlling the hybridization 
between electronic states.  
We study how this can offer another venue for controlled variation
of the magnon dispersion relation and lifetime. 
This latter aspect complements previous theoretical studies 
which have typically examined 
only the adiabatic or infinitely long-lived limit
of collective spin excitations.

This paper is structured as follows. 
We first describe a reference device layout
and introduce the theoretical scheme adopted 
to study from \textit{first principles}
its magnon spectrum
(Sec.~\ref{sec:problem-statement}).
We then present numerical results, for
an Fe monolayer and an Fe bilayer either suspended in vacuum 
or deposited on a Cu substrate.  We show how the magnon 
lifetime and the gap between
low- and high-energy eigenmodes depend on the external electric field
and how this can be traced back to changes of the underlying electronic
structure (Sec.~\ref{sec:numerical-results}).  
We summarize salient aspects of the results in Sec. \ref{sec:discuss} and offer our conclusions in Sec.~\ref{sec:conclusions}.


\section{\label{sec:theory}Computational strategy}
\label{sec:problem-statement}

\begin{figure}[htb]
\centering
\includegraphics[width=8.5cm]{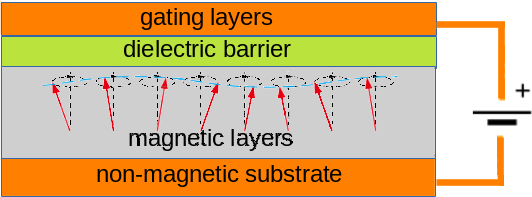}
\caption{\label{fig:device-layout}Schematic device layout. 
Precessing magnetic moments (red arrows) that compose a magnon mode (blue wave)
are studied as a function of an external electric field 
acting along the stacking direction, across a dielectric barrier (green region)
which prevents charge transport.
}
\end{figure}

We consider 
a metallic 2D heterostructure which contains a thin magnetic region 
on top of a non-magnetic substrate  
and which is furthermore capped by a dielectric layer.  
A steady voltage between the substrate 
and an electrode located atop the dielectric barrier
sets up a constant electric field $E_{\text{field}}$ (Fig.~\ref{fig:device-layout}). 
For the sake of clarity and simplicity,  
we model the dielectric barrier by a spacing vacuum gap, 
and we choose respectively Fe and Cu as the material 
of the magnetic and non-magnetic layers.

Our interest lies in how the applied 
voltage
can control 
the spectrum of transverse spin-wave excitations or magnons. 
The magnons are confined within the magnetic layers
because of the negligible proximity-induced spin polarization
in copper.  
However, their dispersion relation $\omega_n(\bm{q})$, 
with $\bm{q}$ being the wave vector confined to the 2D Brillouin zone $\Omega_{BZ}$
and $n$ labeling distinct eigenmodes,
as well as their lifetime, 
depend significantly on the underlying substrate
already in the absence of any applied \eff.

Various dissipation mechanisms can be 
responsible for finite lifetime of magnons
that manifests itself through the $\bm q$-dependent broadening 
of the above dispersion relation $\omega_n(\bm q)$. 
Here we consider a 2D periodic, perfectly long-range ordered (LRO)
scenario in the zero temperature limit, and we neglect therefore
Bloch damping from disorder \cite{Dean1972,Buczek2018}.  
We also neglect dissipation through magnon-magnon scattering \cite{Azevedo2000,Landeros2008,Xue2017}.
On the other hand, we consider Landau damping, 
which is due to the competition between magnons and
single-particle Stoner spin-flip excitations with same energy and momentum,
and which is deemed to be a dominant attenuation mechanism
for magnons propagation in transition metals \cite{Costa2003a}.


\subsection{General approximation strategy
\label{sec:approx-strategy}
}

In the limit of sufficient time-scale separation between fast
electrons and slow precession of atomic magnetic moments, 
we can adopt as starting point the Heisenberg Hamiltonian
\begin{equation}
\label{eq:Heisenberg_Hamiltonian}
H = - \sum_{i \neq j} J_{ij} \hat{\bm e}_i \cdot \hat{\bm e}_j \; ,
\end{equation}
where $\hat{\bm e}_{i}$ is the direction of magnetic
moment around atom at position $\bm{R}_{i}$~\cite{Halilov1998}.
%
%
The exchange coupling parameters $J_{ij}$ can be calculated at a \textit{first principles} 
electronic structure level by employing, for instance, the magnetic force theorem
\cite{Liechtenstein1984,Liechtenstein1987}. 

Extensions of the basic scheme \cite{Udvardi2003,Mankovsky2017a}
can be used to obtain the full tensor
form, $J_{ij}^{\mu\nu}$ with
$\mu(\nu)=x,y,z$,
which can be of particular relevance in connection 
with relativistic effects such as spin-orbit coupling.
Considering for instance ferromagnetic order along $z$,
one can then identify the isotropic exchange interactions of Eq.~\eqref{eq:Heisenberg_Hamiltonian} 
with $J_{ij} = \frac{1}{2} ( J_{ij}^{xx} + J_{ij}^{yy} )$,
and can analogously define a DMI vector \bm $\bm D_{ij} = ( D_{ij}^x, D_{ij}^y, D_{ij}^z )$ with components
$D_{ij}^x = \frac{1}{2} ( J_{ij}^{yz} - J_{ij}^{zy} )$,
$D_{ij}^y = \frac{1}{2} ( J_{ij}^{xz} - J_{ij}^{zx} )$ and
$D_{ij}^z = \frac{1}{2} ( J_{ij}^{xy} - J_{ij}^{yx} )$.
%
%
Liu \ea\ \cite{Liu2011b} discussed how an applied electric field 
can produce an additional DMI term $H_{DM} = \bm D_{ij} \cdot (\bm S_i \times \bm S_j)$, proportional to the perturbation and to the spin-orbit coupling strength.
 
Although reduced dimensionality can have a significant impact on spin-orbit coupling,
magnetism in thin films is known 
to heavily depend on the interplay between substrate and magnetic layers
already at the level of isotropic exchange interactions $J_{ij}$. 
Our goal is to explore 
to what extent the layout of Fig.~\ref{fig:device-layout}
could be used to control magnon spectral features 
by exploiting field-dependent hybridization of electronic states, 
without depending on more subtle relativistic effects.
We remain, therefore, within the description of Eq.~\eqref{eq:Heisenberg_Hamiltonian},
and we neglect other features 
such as magneto-crystalline anisotropy or Gilbert damping \cite{Kunes2002,Udvardi2003,Hickey2009,He2013}.

The precession of atomic magnetic moments around their ground state direction
in the effective magnetic field generated by all their neighbors, 
$\bm B_i^{\textrm{eff}} = \sum_{j\neq i} J_{ij} \hat{\bm e}_j$,
follows the Landau-Lifschitz equation of motion 
and can be studied
as a secular equation problem. 
In particular, the adiabatic magnon spectrum is given by
the eigenvalues of the lattice Fourier-transformed
expression \cite{Halilov1998,Etz2015} 
\begin{equation}
\label{eq:Landau-Lifschitz-secular}
\widehat{N}(\bm{q}) | \omega_n(\bm{q}) \rangle = \omega_n(\bm{q}) |
\omega_n(\bm{q}) \rangle
\;\; ,
\end{equation} 
with explicit matrix elements $\left[ \underline{N}(\bm{q})
\right]_{s,s'} = \langle s | \widehat{N}(\bm{q}) | s' \rangle $.
The subscript $s=1,\ldots,N_{\textrm{sub}}$ labels the (magnetic) sublattices 
with origin $\bm{b}_{s}$.
Each atom lies therefore at position $\bm{R}_{i} = \bm{R}_{I} +
\bm{b}_{s}$, where $\bm{R}_{I}$ is a vector of the periodic lattice.
For a long-range ordered ground state with atomic magnetic moments $\bm m_s = (0,0,m_s^z)$ the
matrix $\underline{N}(\bm{q})$ has elements 
\cite{Pajda2001,Rusz2006,Jacobsson2013,Bergqvist2013}
\begin{equation}
\label{eq:revised-LL-matrix}
\left[ \underline{N}(\bm{q}) \right]_{s,s'}
=
\frac{4}{m_s^z} \Big[ J_{ss'}(\bm{0}) - J_{ss'}(\bm{q}) \Big]
\;\; .
\end{equation}
The Fourier transformation in Eq.~(\ref{eq:Landau-Lifschitz-secular}) 
is performed over all displacements $\bm{R}_{IJ} = \bm
R_I - \bm{R}_J$ between unit cells $I$ and $J$:
\begin{equation}
\label{eq:jij-fourier}
\begin{split}
J_{ss'}(\bm{0}) 
\: =& \:
\delta_{s,s'} 
\sum\limits_{\bm{R}_{IJ}} 
\sum\limits_{s''=1}^{N_{\textrm{sub}}} 
J_{IsJs''} 
\;\; ,
\\
J_{ss'}(\bm{q})
\: =& \:
\sum\limits_{\bm{R}_{IJ}}
J_{IsJs'} \, 
e^{-i \bm{q} \cdot (\bm{R}_{IJ} + \bm b_s - \bm b_{s'}) }
\;\; .
\end{split}
\end{equation}

The above approach towards studying  magnon spectra is intuitive,
computationally expedite, and typically offers good agreement with experiment. However, it does not account for Landau
damping.  
Physically, it originates 
from competition of collective transverse spin-wave excitations
with single-particle spin-flip excitations \cite{Yosida1991,Kubler2000,Kakehashi2012}.  
A comprehensive scheme to account for both collective and single-particle magnetic excitations
is provided by linear response formalism in the framework of the time-dependent
density functional 
theory (TDDFT). 
This  approach focuses on the dynamic transverse  susceptibility
$\underline{\chi}^{+(-)}(\bm{q},\omega)$ 
which describes the response of spin-polarized electrons to a magnetic field
precessing clockwise $(+)$  or anticlockwise $(-)$ 
with the frequency $\omega$.  
This susceptibility is determined by the Dyson-like equation
\begin{equation}
\label{eq:interacting-susceptibility}
\underline{\chi}^{+(-)}(\bm{q},\omega)
=
\left[
\underline{1} -
\underline{\mathring{\chi}}^{+(-)}(\bm{q},\omega)
\underline{f}_{xc}(\bm{q}) 
\right]^{-1}
\underline{\mathring{\chi}}^{+(-)}(\bm{q},\omega)
\;\; , 
\end{equation}
where the kernel $\underline{f}_{xc}(\bm{q})$ is the second derivative 
of the exchange-correlation energy with
respect to local magnetic moment \cite{Katsnelson2004a,Buczek2011}, and
$\underline{\mathring{\chi}}^{+(-)}(\bm{q},\omega)$ is the transverse
susceptibility of non-interacting electrons.
This quantity can be given at the scalar-relativistic level 
in terms of Kohn-Sham eigenstates 
$\phi_{\nu}$ and eigenvalues $\epsilon_{\nu}$ 
solving the spin-polarized Schr\"{o}dinger problem.
Simplifying for a moment the notation through restriction to the $N_{\textrm{sub}}=1$ case, we have \cite{Kubler2000}
\begin{widetext}
\begin{equation}
\label{eq:Kohn-Sham-susceptibility}
\begin{split}
\mathring{\chi}^{+(-)}(\bm{r},\bm{r'},\bm{q},\omega)
 \: = \: 
\lim\limits_{\eta \to 0^+}
\sum\limits_{\nu,\nu'} \int_{\Omega_{BZ}} \dstd \bm{k} \,
&  
\frac{
\phi_{\nu}^{\uparrow(\downarrow),*}(\bm{k},\bm{r}) \,
\phi_{\nu'}^{\downarrow(\uparrow)}(\bm{k}+\bm{q},\bm{r}) \, 
\phi_{\nu'}^{\downarrow(\uparrow),*}(\bm{k}+\bm{q},\bm{r}') \,
\phi_{\nu}^{\uparrow(\downarrow)}(\bm{k},\bm{r}')
}{
\omega \, + \, \istd \eta \, + \, 
\epsilon_{\nu}^{\uparrow(\downarrow)}(\bm{k}) \, - \,
\epsilon_{\nu'}^{\downarrow(\uparrow)}(\bm{k} + \bm{q})
} 
\:
\times 
\\
 & 
\left\{
\theta\left[ E_F - \epsilon_{\nu}^{\uparrow(\downarrow)}(\bm{k}) \right]
\, - \, 
\theta\left[ E_F - \epsilon_{\nu'}^{\downarrow(\uparrow)}(\bm{k} + \bm{q}) \right] 
\right\}
\;\; , 
\end{split}
\end{equation}
\end{widetext}
with the Heaviside step function $\theta(x)=1$ for $x>0$, $\theta(x)=0$ for $x \leq 0$.
The left (right) arrow selects the spin polarization
relevant for the clockwise (anticlockwise) precession of the moments
in response to the infinitesimal perturbation of the rotating magnetic field.  
The wave vectors for $\bm{k}$, $\bm{k} +\bm{q}$
are considered within the Brillouin zone $\Omega_{BZ}$, 
and the positions $\bm{r}$, $\bm{r}'$ are restricted to the
Wigner-Seitz cells around sites $\bm{R}_I,\bm{R}_J$, respectively.
The quantities in Eqs.~\eqref{eq:interacting-susceptibility} and \eqref{eq:Kohn-Sham-susceptibility} 
can be cast in matrix form 
by adopting, e.g., a combined basis set of spherical harmonics and
orthogonal polynomials to represent the $\bm r$, $\bm{r}'$ dependence \cite{Staunton2000,Buczek2011}.

Thanks to the fluctuation-dissipation theorem \cite{Kubo1957},
the propensity of a material to host a magnetic excitation
with wave vector $\bm{q}$ and energy $\omega$ is marked by large values 
in the loss matrix
$\Im \underline{\chi}^{+(-)} (\bm{q},\omega)$.
Technically, this is due to zeros from the first term,
$\underline{1} -
\underline{\mathring{\chi}}^{+(-)}(\bm{q},\omega)
\underline{f}_{xc}(\bm{q})$,
as well as to singularities from the second term, $\underline{\mathring{\chi}}^{+(-)}(\bm{q},\omega)$, in Eq.~\eqref{eq:interacting-susceptibility}.
The outcome can be studied 
by examining the eigenvalues of $\Im \underline{\chi}^{+(-)} (\bm{q},\omega)$
as a function of $\bm q$ and $\omega$
\cite{Antropov2003,Buczek2011}.

Long-living collective excitations (magnons) are characterized 
by the occurence, at each energy and wave-vector,
of as many sharply defined eigenvalues
as the number of magnetic sublattices in the unit cell \cite{Buczek2011}.
By following the sequence of such
peaks one can reconstruct their dispersion relation
and compare it for instance with the simpler $\omega_n(\bm q)$ outcome from Eq.~\eqref{eq:Landau-Lifschitz-secular}.
 
Landau damping instead manifests itself through the emergence 
of multiple, 
no longer well-separated eigenvalues
which lead in practice to a broadened magnon dispersion.
The broadening can be interpreted as inversely proportional 
to finite magnon lifetime due to competition with Stoner single-particle excitations.
These spin-flip transitions are described in particular by
the non-interacting susceptibility $\mathring{\chi}^{+(-)}(\bm{r},\bm{r'},\bm{q},\omega)$ \cite{Buczek2011}
and are entirely neglected in the secular equation problem of Eq.~\eqref{eq:Landau-Lifschitz-secular}.

In order to approximately account for this aspect of the magnon physics,
we apply here at a \textit{first principles} level
an approximative procedure 
that has been proposed, among others, by Yosida \cite{Yosida1991} for simplified theoretical models,
and adopted, e.g., by Kirschner et al.~\cite{Kirschner1986,Venus1988,Vollmer2003}
for the interpretation of spin-polarized electron energy loss experiments 
in metallic thin films.

The procedure consists of two steps.
First we obtain the adiabatic dispersion relation $\omega_n(\bm{q})$
from Eq.~\eqref{eq:Landau-Lifschitz-secular}.
This involves diagonalizing for each $\bm q$ 
the real $N_{\textrm{sub}} \times N_{\textrm{sub}}$ matrix
defined in Eq.~\eqref{eq:revised-LL-matrix}.
Such a procedure is much simpler than dealing with complex matrices of Eqs.~\eqref{eq:interacting-susceptibility} and \eqref{eq:Kohn-Sham-susceptibility},
which need to be dealt with not only for each $\bm{q}$ but also for
every trial energy $\omega$ and which are also much bigger,
depending on the sampling in $\bm r$ and $\bm{r}'$.

Subsequently, 
the intensity of single-particle excitations
$S^{+(-)}_n(\bm{q})$ is obtained by considering 
only Stoner spin-flip 
transitions between occupied 
and unoccupied Kohn-Sham states, 
such that their difference in energy and momentum corresponds to
the magnon eigenmode under consideration $|\omega_n(\bm q) \rangle$.
%
The number of relevant transitions is estimated 
by convoluting the spin-polarized electronic Bloch spectral functions
$A^{\uparrow(\downarrow)}(\bm{k},s,E) = -\frac{1}{\pi} \Im \,
G^{\uparrow(\downarrow)}(\bm{k}, s,E)$
where the electronic Green's function $G^{\uparrow(\downarrow)}(\bm{k}, s,E)$ is the Lehmann resummation of Kohn-Sham eigenstates and eigenvalues
already appearing in Eq.~\eqref{eq:Kohn-Sham-susceptibility}.
In practice we adopt the KKR construction 
to directly obtain these Green functions \cite{Ebert2011}, 
calculate the Heisenberg exchange parameters $J_{ij}$ \cite{Liechtenstein1987}
and solve the secular equation problem of Eq.~\eqref{eq:Landau-Lifschitz-secular}, 
and then we evaluate the expression
%
\begin{widetext}
\begin{equation}
\label{eq:Stoner-convolution}
\begin{split}
S^{+(-)}_n(\bm{q})
=&
\int_{E_{\textrm{min}}}^{E_{\textrm{max}}} \dstd E 
\int_{\Omega_{BZ}} \dstd^3 k
\sum\limits_{s=1}^{N_{\textrm{sub}}}
A^{\uparrow(\downarrow)}(\bm{k},s,E) \, \theta(E_{F}-E) \: 
A^{\downarrow(\uparrow)}(\bm{k}+\bm{q},s,E + \omega_n(\bm{q})) \,
\theta(E + \omega_n(\bm{q}) - E_{F})  \,
\times
\\
& \times \sqrt{ \Re[v_{n,s}(\bm{q})]^2 + \Im[v_{n,s}(\bm{q})]^2 },
\end{split}
\end{equation}
\end{widetext}
%
where the double integration samples the full Brillouin zone $\Omega_{BZ}$ 
and the energy interval 
$E_{\textrm{min}} = E_F - \max[ \omega_n(\bm{q})]$,
$E_{\textrm{max}} = E_F + \max[ \omega_n(\bm{q})]$
around the Fermi level
$E_F$.
Occupied and unoccupied states are selected via the Heaviside step function, similarly to Eq.~\eqref{eq:Kohn-Sham-susceptibility}.
Finally, the last term in Eq.~\eqref{eq:Stoner-convolution}
is the sublattice-projected magnitude of the complex-valued eigenvector 
$ | \omega_n(\bm q) \rangle := ( v_{n,1}(\bm q),v_{n,2}(\bm q),\ldots, v_{n,N_{\textrm{sub}}}(\bm q) )^{\dagger}$
from Eq.~\eqref{eq:Landau-Lifschitz-secular}. In general,
this quantity describes 
how the $n$ magnon mode involves deviations from the ground state 
at each magnetic sublattice \cite{Halilov1998}. 
In this context, it is used to perform a weighted sum
of Stoner spin-flip transitions which also originate from that sublattice,
and which are assumed to compete proportionally more 
with the specific magnon mode, 
depending on how it involves the same atoms.

Compared to Eq.~\eqref{eq:Kohn-Sham-susceptibility}, 
the energy and momentum convolution of Eq.~\eqref{eq:Stoner-convolution}
only involves real quantities.
We use the result to produce a magnon spectral function 
which includes the finite lifetime 
\begin{equation}
\label{eq:approximated_magnon_spectral_function}
A_{\textrm{mag}}(\bm{q},n,\omega) \: = \: 
- \lim_{\eta \to 0^{+}} 
\frac{ |\omega_n(\bm{q})\rangle \: \langle \omega_n(\bm{q})| }
{ \omega \, + \, 
  \istd [ \eta \, + \, S_n^{+(-)}(\bm{q}) ] - \, \omega_n(\bm{q}) }
\;\; .
\end{equation}

We note that the approach is not as robust as the more rigorous but demanding formulation 
in terms of the loss matrix $\Im \underline{\chi}^{+(-)} (\bm{q},\omega)$ from Eq.~\eqref{eq:interacting-susceptibility}. 
Among various simplifications behind it, 
we deem as most severe the separate evaluation 
of the adiabatic dispersion $\omega_n(\bm q)$ and of the broadening function $S^{+(-)}_n(\bm{q})$. 
These quantities are used within Eq.~\eqref{eq:approximated_magnon_spectral_function} 
to approximate complex magnon poles 
which would, in an exact treatment, follow from analyzing 
the dynamic transverse susceptibility. 

The TDDFT Eq.~\eqref{eq:interacting-susceptibility} construction of the magnon spectral function 
evaluates collective and single-particle spin-flip excitations on equal footing, meaning that their relative spectral weights gets redistributed, 
depending for instance on the location of the wave vector $\bm q$ within the Brillouin zone, but it remains on the whole conserved.
The approximated construction of Eq.~\eqref{eq:approximated_magnon_spectral_function} reproduces 
some of the same features, 
but does not guarantee 
conservation of the total spectral weight \cite{Edwards1978,Buczek2011}. 

%
%
However, our aim is not to obtain absolute values for the Landau damping
but rather to investigate its relative changes as a function 
of the externally applied electric field efficiently.  
As long as the inaccuracies of the more expedite but less robust approach
depend only weakly on this perturbation, we can expect reasonable trends 
for the ratio between lifetime estimated with $E_{\textrm{field}}=0$ and $E_{\textrm{field}} \neq 0$.


\subsection{Finite electric field and other technical aspects}

The results discussed in the following have been produced using the {\em ab~initio}
spin-polarized multiple-scattering or Korringa-Kohn-Rostoker (KKR)
Green function formalism \cite{Ebert2011} as implemented in the {\sc
  SPRKKR} code \cite{sprkkr2022}.  
The self-consistent field (SCF) ground state for the 2D heterostructure of Fig.~\ref{fig:device-layout} was obtained by solving the DFT problem in fully relativistic mode, relying on the local spin density
approximation (LSDA) with the Vosko, Wilk and Nusair parametrisation
for the exchange and correlation term \cite{Vosko1980}.  

To deal with systems with only 2D periodicity, we used the
tight-binding or screened KKR method \cite{Zeller1995}.  Fe monolayers and
bilayers suspended in vacuum were modeled by slabs consisting of one
or two Fe layers embedded in vacuum represented by four layers
of empty sites at each site.  Fe monolayers or bilayers deposited on
Cu(001) were treated as truly semi-infinite systems: the electronic
structure was reconverged within the topmost eleven or ten substrate
layers, while at the bottom of this interaction zone the electronic
structure was matched to the bulk.  
For all our systems we used experimental unit cell parameters of bulk copper, neglecting lattice relaxations, and assuming out-of-plane easy axis of
magnetization \cite{Allenspach1992,Vaz2008a}.  The geometry of Fe
layers suspended in vacuum is taken the same as the geometry of the
layers deposited on Cu(001).

The external electric field is introduced similarly as in
Refs.~\cite{Simon2021,Mankovsky2021a}, namely, 
by considering above the Fe layers an auxiliary array of point charges, separated from the surface by vacuum, during calculation of the SCF solutions and all other quantities.
For sufficient areal density and vertical separation,
this layer generates an electric field which can
be considered constant \cite{Zhang2009d,Ignatiev2011},
with intensity
\begin{equation}
\label{eq:Efield}
E_{\textrm{field}} \: = \: \frac{ Q_{\textrm{aux}} }{ 2 \epsilon_0 A }
\;\; ,
\end{equation}
where $Q_{\textrm{aux}}$ is the point charge (positive 
for a field oriented antiparallel to the surface normal $\widehat{z}$)
per area of the 2D unit cell $A$, and  $\epsilon_0$ is the vacuum permitivity.

For the multipole expansion of the Green function, the angular momentum
cutoff $\ell_{\text{max}}=3$ was used.  
The energy integrals to obtain the SCF-DFT solutions, 
as well as the isotropic Heisenberg exchange interactions 
from the magnetic force theorem \cite{Liechtenstein1987},
were evaluated by contour integration on a semicircular path within
the complex energy plane using 32 Gaussian-Legendre abscissae.
The Brillouin zone integrals used an equispaced mesh with 16000 $\bm k$-points
or more, over the whole $\Omega_{BZ}$. 
The Stoner expression Eq.~\eqref{eq:Stoner-convolution} was evaluated by sampling energy points parallel and near to the real axis.

For the ferromagnetic ground states studied in Sec.~\ref{sec:numerical-results}
we only need to consider one chirality,
meaning that we restrict ourselves to the $(+)$ variant 
of Eqs.~\eqref{eq:interacting-susceptibility}-\eqref{eq:Stoner-convolution}
\cite{Yosida1991,Kakehashi2012,Buczek2011}.


\section{\label{sec:numerical-results}Results}

We discuss here results for a Fe monolayer and a Fe bilayer,
both suspended in vacuum as well as deposited on Cu(001) surface.


\subsection{\label{sec:suspend-mono-bilay}Fe monolayer and Fe bilayer
  in vacuum}

\begin{figure}[htb]
\centering
\includegraphics[width=9.5cm]{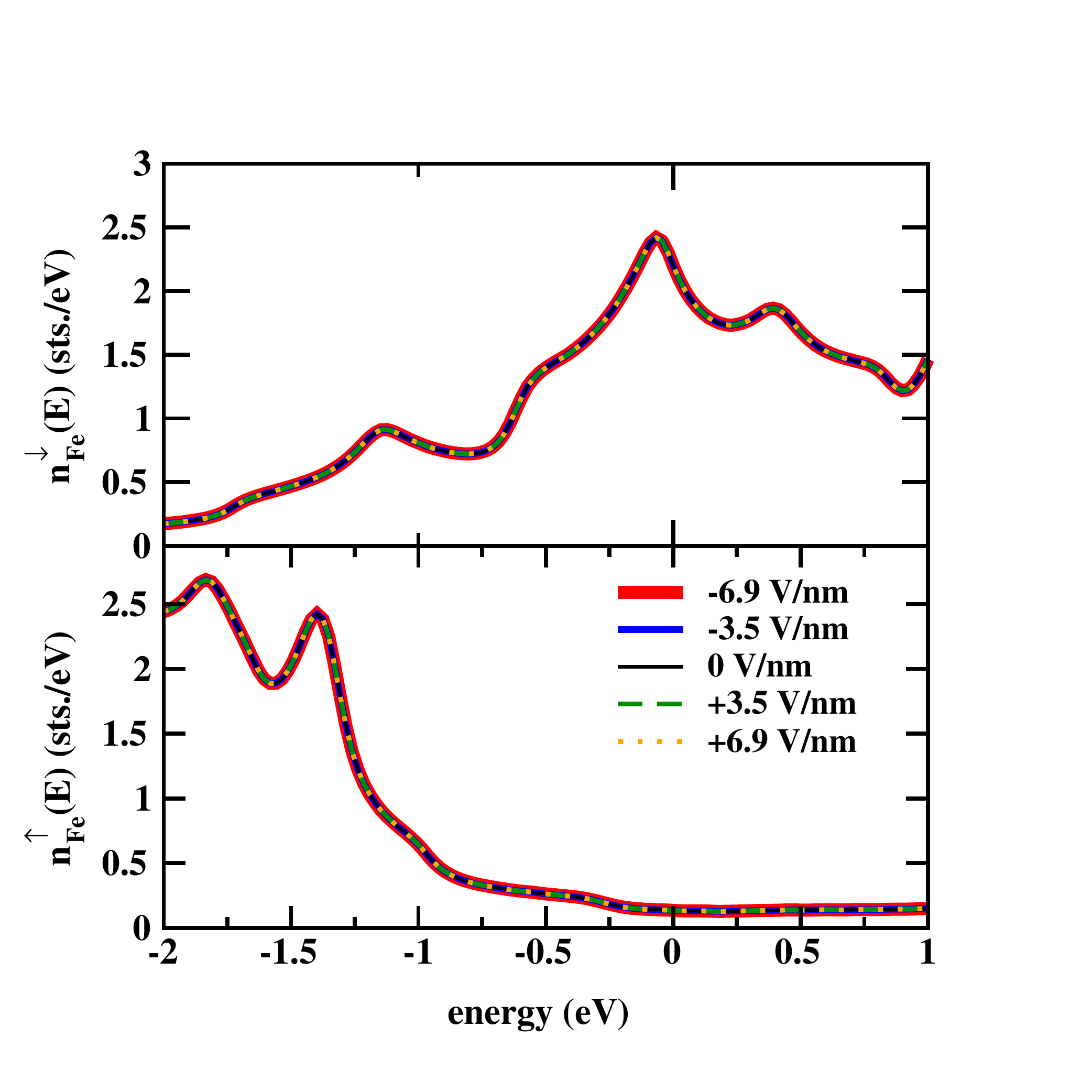}
\caption{\label{fig:DOS-QCHRLAY-Fe-mono}DOS of a Fe monolayer
suspended in vacuum for different values of \eff.  
All the curves fall essentially on top of each other, with no discernible
effects from the electric field.
}
\end{figure}

\begin{figure}[htb]
 \centering
\includegraphics[width=9.5cm]{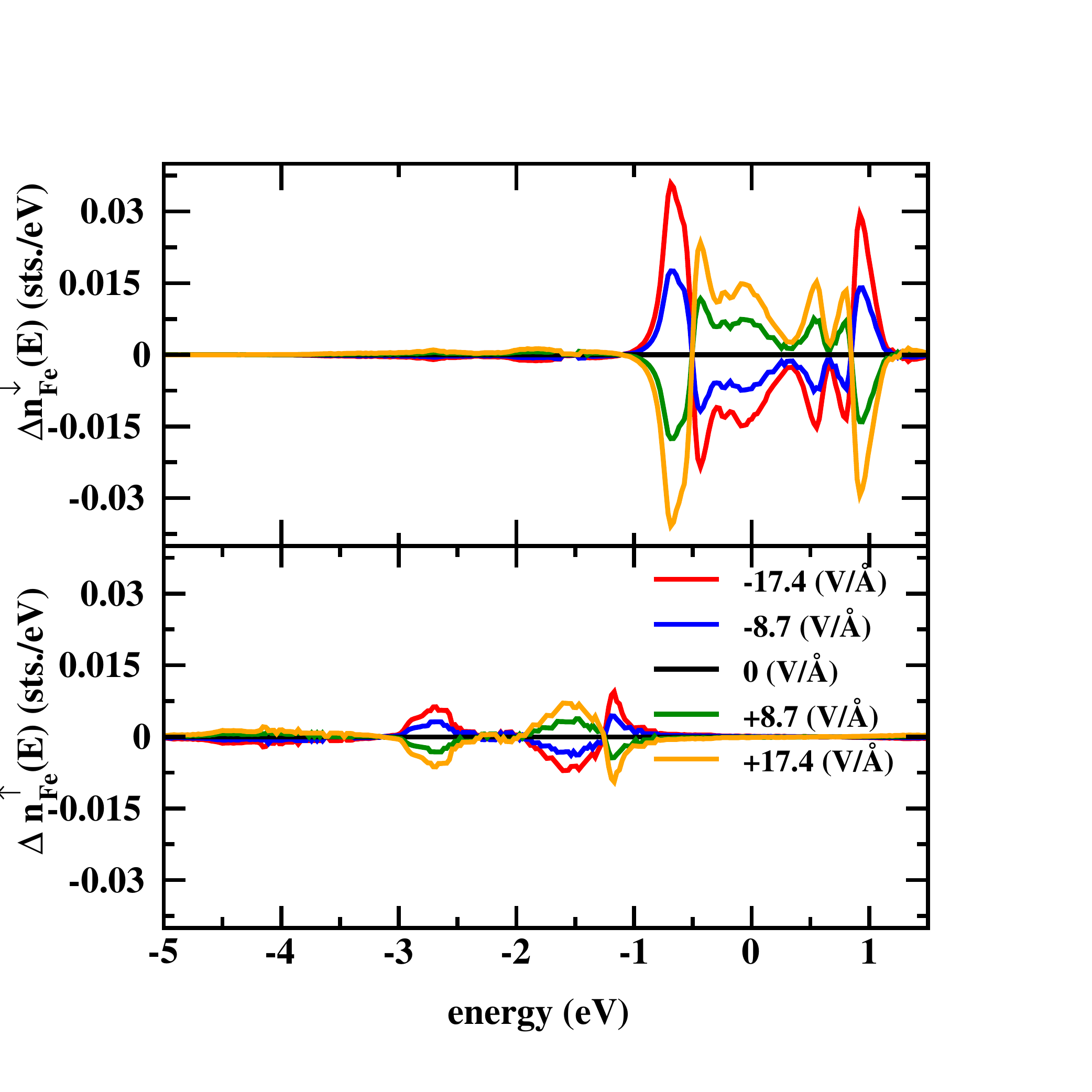}
\caption{\label{fig:delta-DOS-QCHRLAY-Fe-bilay}Difference between the DOS 
projected on individual layers of a Fe bilayer as a function of E$_{\textrm{field}}$.}
\end{figure}

We begin examining how the external electric field influences
the spin-polarized density of states (DOS).  Results for a Fe
monolayer are shown in Fig.~\ref{fig:DOS-QCHRLAY-Fe-mono}, 
with no visible effects. 
Magnon spectra appear similarly robust with respect to the perturbation
and are therefore not shown.

If a second iron sheet is added, changes in the layer-resolved DOS 
start to appear but they are
still very small. Therefore, to highlight the influence of the
external perturbation E$_{\textrm{field}}$, we consider the difference 
between the DOS projected on individual layers,
\[
\Delta n^{\uparrow(\downarrow)}(E) \: = \: 
n^{\uparrow(\downarrow)}_{\text{Fe}_{1}}(E) \, - \,
n^{\uparrow(\downarrow)}_{\text{Fe}_{2}}(E)
\; .
\]
%
%
The outcome is shown in Fig.~\ref{fig:delta-DOS-QCHRLAY-Fe-bilay}.  
If there is no external field, this difference is obviously zero 
because the bilayer is symmetric.  
With a finite \eff, the symmetry is removed and small energy- and
spin-dependent transfer of electronic states between both layers occurs.
This transfer is more pronounced for the minority states.
Swapping the polarity of the perturbation, or the labeling of Fe$_1$ and Fe$_2$ layers,
is equivalent to the $z \to -z$ coordinate transformation and
leads to identical results. This will only change in the presence 
of a substrate which lifts the symmetry, as discussed in Sec.~\ref{sec:mono-Cu001} below.

\begin{figure}[htb]
 \centering
\includegraphics[width=8.5cm]{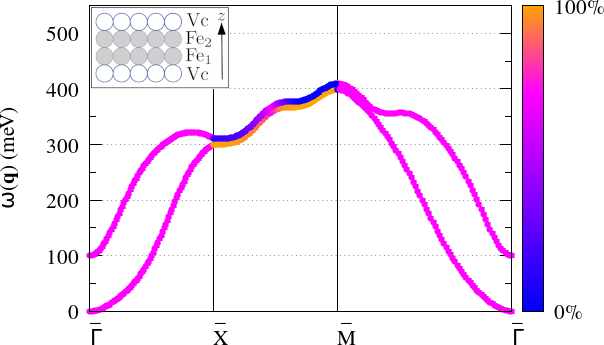}
\caption{\label{fig:suspend-bilay-eigenmodes}Adiabatic magnon
spectrum for the Fe bilayer suspended in vacuum with $E_{\textrm{field}}=0$.  
The $\omega_2(\bm q)$ solution is plotted 
with an artificial offset of +10 meV, 
to allow visualization where energy degenerate.
The color coding represents the magnitude 
of the corresponding complex eigenvectors,
projected on the Fe$_2$ layer.
}
\end{figure}


With only two magnetic layers, the secular equation problem expressed by
Eqs.~\eqref{eq:Landau-Lifschitz-secular} and \eqref{eq:revised-LL-matrix}
reduces to
diagonalizing the matrix
\begin{equation}
\label{eq:Landau-Lifschitz-bilayer}
\underline{N}(\bm{q}) 
\hspace{-.05cm} = \hspace{-.05cm}
4 \hspace{-.05cm} \sum_{\bm{R}_{IJ}} \hspace{-.075cm} 
\left(\begin{array}{cc}
\hspace{-.075cm} 
\frac{ J_{IJ}^{11} + J_{IJ}^{12} - J_{IJ}^{11} e^{-i \bm{q} \cdot \bm{R}_{IJ} } }
{ m_1^z } & 
\hspace{-.075cm} 
\frac{ - J_{IJ}^{12} e^{-i \bm{q} \cdot (\bm{R}_{IJ} + \bm b_1 - \bm b_2)} }
{ m_1^z } 
\hspace{-.075cm}
\\[1ex]
\hspace{-.075cm} 
\frac{ - J_{IJ}^{21} e^{-i \bm{q} \cdot (\bm{R}_{IJ} + \bm b_2 - \bm b_1)} }
{ m_2^z }  & 
\hspace{-.075cm} 
\frac{ J_{IJ}^{21} + J_{IJ}^{22} - J_{IJ}^{22} e^{-i \bm{q} \cdot \bm{R}_{IJ} } }
{ m_2^z } \hspace{-.075cm}\\
\end{array}\right)
\end{equation}
Results are shown in Fig.~\ref{fig:suspend-bilay-eigenmodes}.  We observe that
eigenvalues are distinct between the
$\overline{\Gamma}$ and the $\overline{X}$ point and between the
$\overline{M}$ and the $\overline{\Gamma}$ point, i.e., when
going from the center of the 2D Brillouin zone to its corners.
For these portions of the spectrum, magnetic precession involves atoms from both layers.
On the contrary, along the $\overline{X}$--$\overline{M}$ segment,
i.e., at the Brillouin zone edge,
eigenvalues are degenerate
but precession involves exclusively one or the other iron sheet.

\begin{figure}[htb]
 \centering
\includegraphics[width=8.5cm]{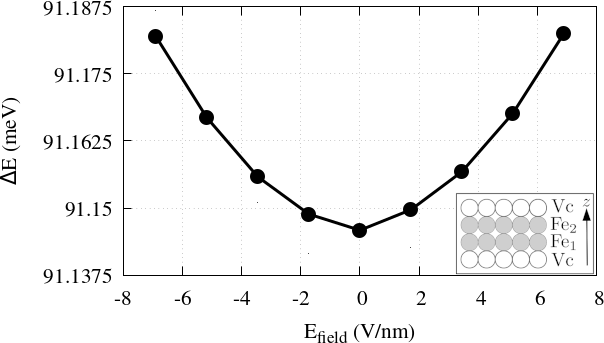}
\caption{\label{fig:gap-bilay-suspended}Energy gap between the high- and
  low-energy magnon branches at $\bm{q} = \overline{\Gamma}$ for an iron
  bilayer suspended in vacuum
  (cf.~Fig.~\protect\ref{fig:suspend-bilay-eigenmodes}) evaluated as a
  function of E$_{\textrm{field}}$.  }
\end{figure}

The effect of the external electric field on the magnon spectra
is again very weak for this suspended Fe bilayer, so that it would
be hardly visible in a plot. Therefore we focus just on the gap
between the high- and low-energy branches at the
$\overline{\Gamma}$ point (see
Fig.~\ref{fig:suspend-bilay-eigenmodes}).
This gap can be
evaluated as
\[
\Delta E \: = \: 
\omega_2(\overline{\Gamma}) - \omega_1(\overline{\Gamma}) 
\: = \:
4 \sum_{\bm{R}_{IJ}} 
J_{IJ}^{12} 
\,
\frac{ m_1^z + m_2^z }{ m_1^z \, m_2^z }
\;\; .
\]
The dependence of this gap on E$_{\textrm{field}}$ is shown in
Fig.~\ref{fig:gap-bilay-suspended}.  We observe 
a very small variation for the considered range of E$_{\textrm{field}}$, just about 0.05~\%. 
Similarly as for
Fig.~\ref{fig:delta-DOS-QCHRLAY-Fe-bilay}, the graph in
Fig.~\ref{fig:gap-bilay-suspended} is symmetric with respect to the polarity of
the external field, in accordance with the interchangeable role of
layer~1 and layer~2 in the absence of a substrate.


\subsection{\label{sec:mono-Cu001}Fe monolayer on Cu(001) substrate}

\begin{figure}[htb]
\centering
\includegraphics[width=9.5cm]{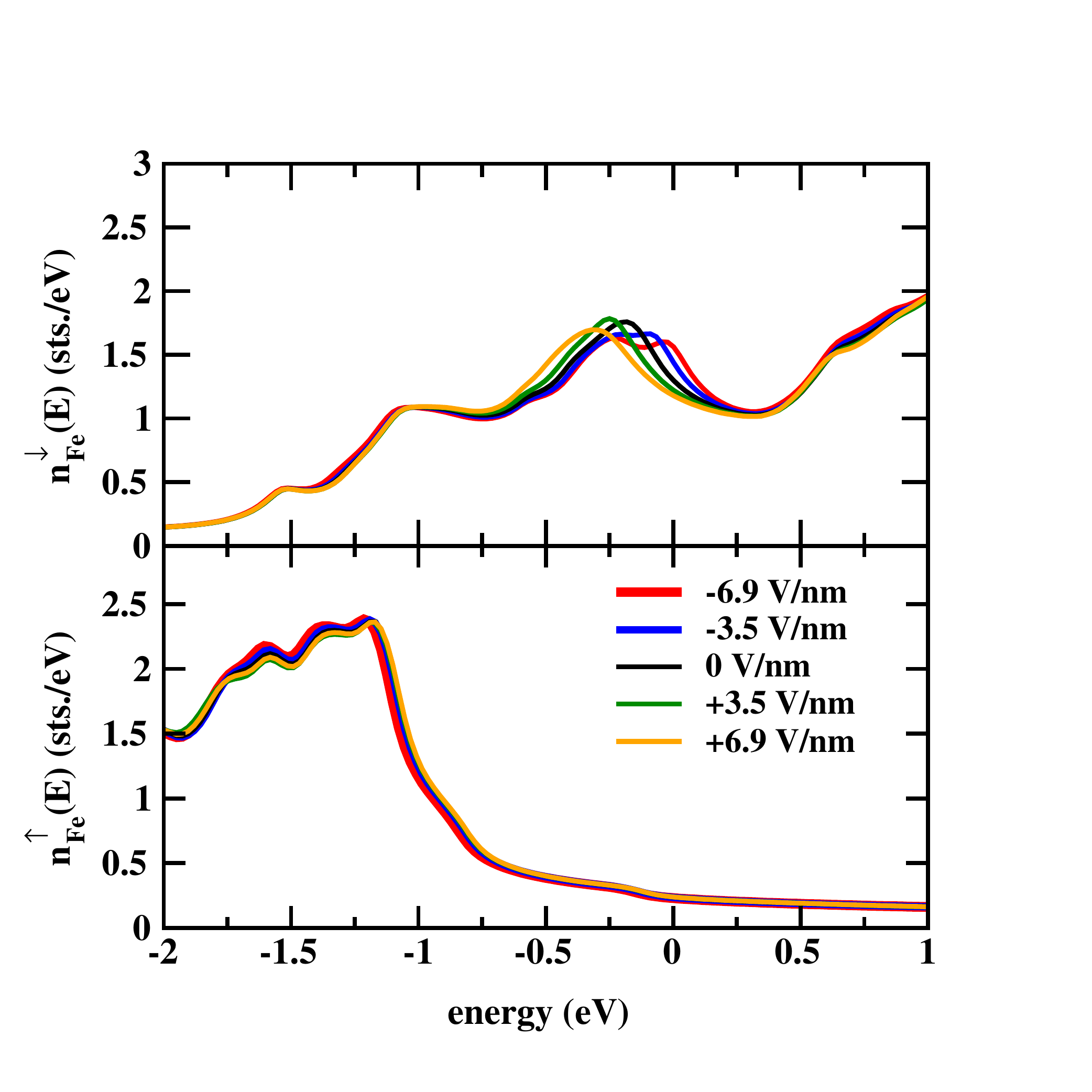}
\caption{\label{fig:DOS-QCHRLAY-CuFe001}Spin-polarized Fe-projected
  DOS for a Fe monolayer on Cu(001) for different intensities and
  polarities of the external electric field.  }
\end{figure}

\begin{figure}[htb]
\centering
\includegraphics[width=8.cm]{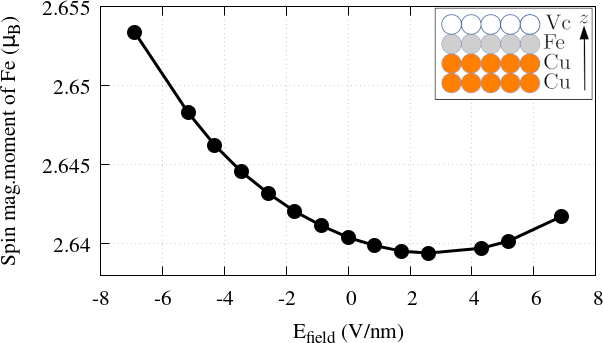}
\caption{\label{fig:SMT-QCHRLAY-CuFe001}Dependence of the magnetic
  moments at Fe sites on the external electric field for a Fe
  monolayer on Cu(001).  
  }
\end{figure}

Larger effects can be expected for supported iron sheets,
because here the asymmetry introduced by the external field couples
with the asymmetry stemming from the substrate.
Fig.~\ref{fig:DOS-QCHRLAY-CuFe001} shows how the spin-polarized Fe-projected DOS 
varies with \eff\ for a Fe monolayer on Cu(001).  The
changes are now clearly visible, contrary to the
situation for layers suspended in vacuum investigated in
Figs.~\ref{fig:DOS-QCHRLAY-Fe-mono} and
\ref{fig:delta-DOS-QCHRLAY-Fe-bilay}.

The corresponding change of the magnetic
moment with E$_{\textrm{field}}$ is shown in
Fig.~\ref{fig:SMT-QCHRLAY-CuFe001}.  The presence of the substrate
means that the polarity of the external electric field matters this
time --- unlike in the case of suspended layers,
as evidenced e.g. in Fig.~\ref{fig:gap-bilay-suspended}.  
Overall, the variation in the magnetic moment is quite small,
about 0.5~\%.  
\begin{figure}
\centering
\begin{tabular}{cc}
\rotatebox{90}{\hspace{1.65cm}E$_{\textrm{field}}$= -5.2 V/nm} 
\includegraphics[width=8.cm]{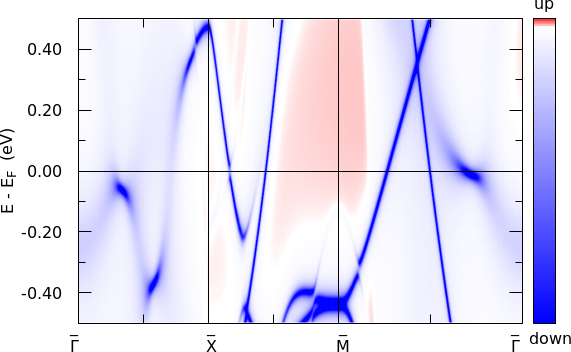}
\\
\rotatebox{90}{\hspace{1.65cm}E$_{\textrm{field}}$= 0 V/nm} 
\includegraphics[width=8.cm]{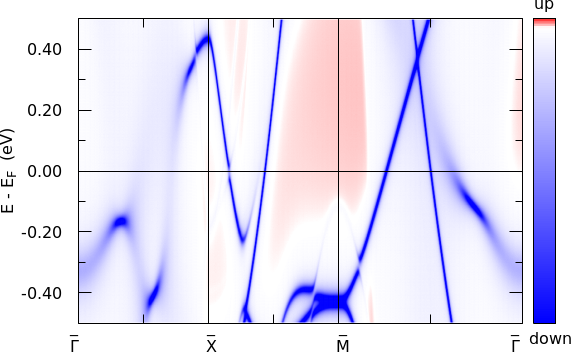}
\\
\rotatebox{90}{\hspace{1.65cm}E$_{\textrm{field}}$= +5.2 V/nm} 
\includegraphics[width=8.cm]{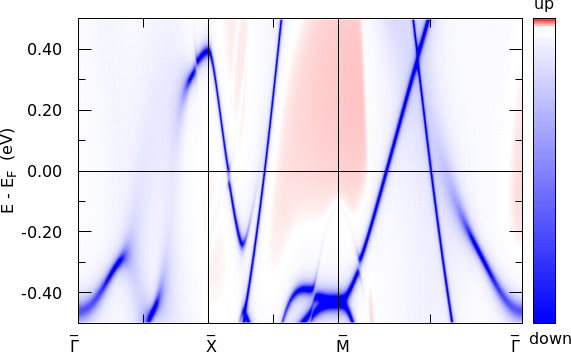}
\\
\end{tabular}
\caption{\label{fig:BSF-Fe-mono-Cu001}Fe-projected Bloch spectral
  function for a Fe monolayer on Cu(001), color-coded to indicate
  the predominantly down spin-polarization of electronic states
  at the Fermi level. 
  From top to bottom: results for
  E$_{\textrm{field}}$= -5.2, 0, or +5.2 (V/nm).  }
\end{figure}
 
A more detailed view
can be obtained by inspecting the projection of the Bloch spectral
function at the Fe site.  Its dependence on \eff\ is outlined in
Fig.~\ref{fig:BSF-Fe-mono-Cu001}.
We show an interval around the Fermi level,
which corresponds to the $\max[\omega_n(\bm q)]=0.5$ eV energy range
of magnons in iron thin films.

Note that the Bloch spectral
function exhibits the characteristic broadening from lack of
periodicity along the $z$ direction. Even though the general look of
all three graphs is the same in Fig.~\ref{fig:BSF-Fe-mono-Cu001}, a
systematic dependence of the position of certain features on \eff\ is
evident: for example, the energy positions of the local maximum within
0.3~eV below $E_{F}$ for $\bm{k}$ between $\overline{\Gamma}$ and
$\overline{X}$ or the energy positions of the inflection point within
0.3~eV below $E_{F}$ for $\bm{k}$ between $\overline{M}$ and
$\overline{\Gamma}$. 
%

\begin{figure}[htb]
\centering
%
\includegraphics[width=8.5cm]{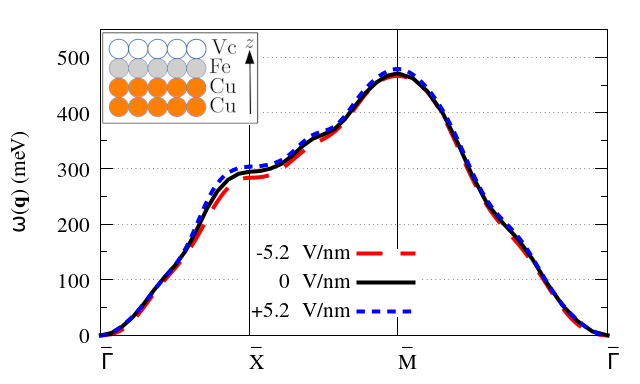}
\caption{\label{fig:magnons-QCHRLAY-mono-Cu001}Adiabatic magnon
  spectrum of a Fe monolayer on Cu(001) for selected values of
  E$_{\textrm{field}}$= -5.2, 0, and +5.2
  (V/nm). 
}
\end{figure}

\begin{figure}[htb]
\centering
%
%
\begin{tabular}{c}
\includegraphics[width=8.cm]{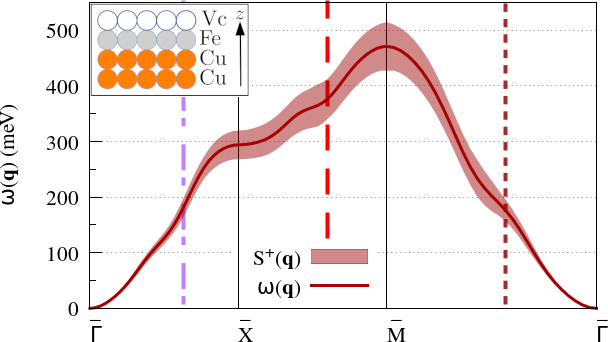} \\
\includegraphics[width=8.5cm]{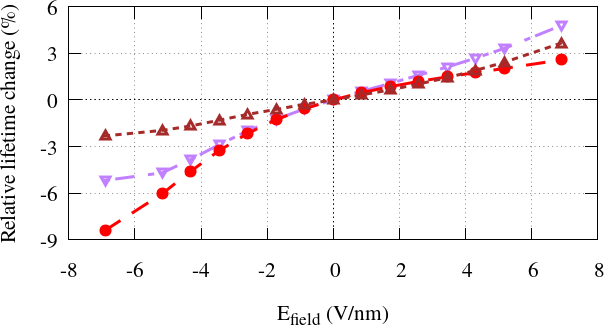} \\
\end{tabular}
\caption{\label{fig:damping-QCHRLAY-38-CuFe}Top panel: Magnon
  spectrum for a Fe monolayer on Cu(001) for $E_{\textrm{field}}=0$,
  depicting eigenvalues according to
  Eq.~\eqref{eq:Landau-Lifschitz-secular} (darker line) together with
  the corresponding intensity of Stoner excitations obtained by
  evaluating Eq.~\eqref{eq:Stoner-convolution} (lighter shaded area,
  in arbitrary units).  Bottom panel: Relative change of the magnon
  lifetime (obtained as the inverse of the Stoner intensity) with
  \eff, for three choices of the $\bm{q}$-vector indicated in the top
  graph by differently dashed vertical lines of matching colors.
    }
\end{figure}

We show in Fig.~\ref{fig:magnons-QCHRLAY-mono-Cu001}
the dispersion relation $\omega(\bm q)$ obtained 
according to Eq.~\eqref{eq:Landau-Lifschitz-secular} 
for the same three values of \eff\
considered in Fig.~\ref{fig:BSF-Fe-mono-Cu001}. 
We observe a very limited dependence.
However, the situation is different for the Stoner spectrum estimated
by means of Eq.~\eqref{eq:Stoner-convolution}.  
Results for \eff=0 are first illustrated 
in the top graph of Fig.~\ref{fig:damping-QCHRLAY-38-CuFe} as a broadening 
of the dispersion $\omega(\bm{q})$. 
The qualitative outcome of increasing Landau damping 
as we move away from the $\overline{\Gamma}$ point 
compares well both with experiments and with more comprehensive TDDFT calculations \cite{Buczek2011}.
We interpret this broadening as inversely proportional to the magnon lifetime.
The bottom graph of Fig.~\ref{fig:damping-QCHRLAY-38-CuFe} shows the 
relative change of this quantity with \eff. Results are depicted for three choices of the $\bm{q}$-vector, indicated by dashed lines in
the top graph of the same figure.
It is evident that varying E$_{\textrm{field}}$ leads to significant
changes in the Stoner spectrum and, consequently, to different magnon lifetime.
The general trend is that a positive \eff\ decreases the Landau damping thereby extending the magnon lifetime, whereas a negative \eff\
increases the damping and therefore reduces the magnon
lifetime.  The effect of a negative \eff, generated by having negative
point charges above the Fe/Cu(001) semi-infinite system, appears to be
larger than the effect of a positive \eff.


\subsection{\label{sec:bilay-Cu001}Fe bilayer on Cu(001)}

\begin{figure}[htb]
\centering
\includegraphics[width=8.5cm]{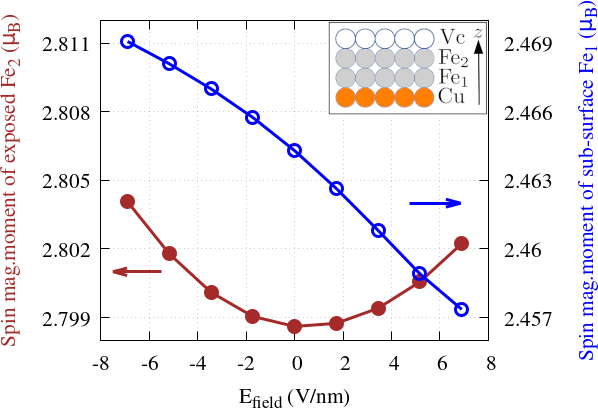}
\caption{\label{fig:SMT-QCHRLAY-bilay-Cu001}Spin magnetic moment
  vs. E$_{\textrm{field}}$ for the exposed Fe$_2$ (brown full circles,
  left scale) and subsurface Fe$_1$ (blue empty circles, right scale)
  for an iron bilayer over Cu(001) substrate.}
\end{figure}

In the previous part Sec.~\ref{sec:mono-Cu001} we investigated a
system with a single magnon eigenmode.  In order to have more
eigenmodes, it is necessary to consider more than a single Fe sheet.
The Cu substrate has only a negligible induced magnetic moment
and thus cannot host magnons.  We consider in this part an iron
bilayer on Cu(001), again assuming out-of-plane easy axis of
magnetization and the same unrelaxed lattice parameters as in the previous
sections, to facilitate comparison.

We first examine the dependence of the magnetic moments in both Fe layers on E$_{\textrm{field}}$.  For the upper Fe$_2$ layer, exposed to
the vacuum, this dependence has got a similar nonmonotonous profile as
for the iron monolayer on Cu(001) (compare the line with full
circles in Fig.~\ref{fig:SMT-QCHRLAY-bilay-Cu001} with
Fig.~\ref{fig:SMT-QCHRLAY-CuFe001}).  On the other hand, the magnetic
moments decrease almost linearly with increasing \eff\ 
for the subsurface Fe$_1$ layer (blue line with empty circles
in Fig.~\ref{fig:SMT-QCHRLAY-bilay-Cu001}).  The total change of the
magnetic moment across the investigated range of \eff\ is about
0.5~\% for both layers, similarly as in the case of a Fe monolayer on
Cu(001). 

\begin{figure}[htb]
 \centering
\includegraphics[width=8.5cm]{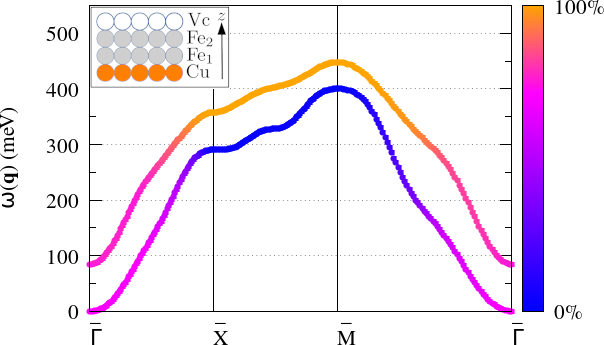}
\caption{\label{fig:CuFe-bilay-eigenmodes}Adiabatic magnon
spectrum for a Fe bilayer on Cu(001) and with E$_{\textrm{field}}=0$.
The color coding represents the magnitude 
of the corresponding complex eigenvectors,
projected on the Fe$_2$ layer
(as in Fig.~\ref{fig:suspend-bilay-eigenmodes}).}
\end{figure}

The adiabatic magnon dispersion is shown in Fig.~\ref{fig:CuFe-bilay-eigenmodes}. 
Some qualitative differences appear with respect to the case of a Fe bilayer suspended in vacuum.  
In particular, the substrate removes the energy degeneracy 
also for $\bm{q}$
points along the $\overline{X}$--$\overline{M}$ path.
On the other hand, the suspended bilayer and the bilayer deposited on
Cu(001) exhibit alike involvement of
individual iron sheets' moments in hosting the magnons. 
The two eigenmodes involve precession of magnetic moments equally from both iron sheets near to
$\overline{\Gamma}$, and from only one or the other layer 
away from the origin of the Brillouin zone.
The high-energy branch involves only the
subsurface Fe$_1$ atoms 
along the $\overline{X}$--$\overline{M}$ path, 
whereas the low-energy branch 
involves only the surface Fe$_2$ atoms. 
A similar $\bm q$-resolved decomposition can be observed 
for the suspendend bilayer of Fig.~\ref{fig:suspend-bilay-eigenmodes}.

\begin{figure}[htb]
 \centering
\includegraphics[width=8.5cm]{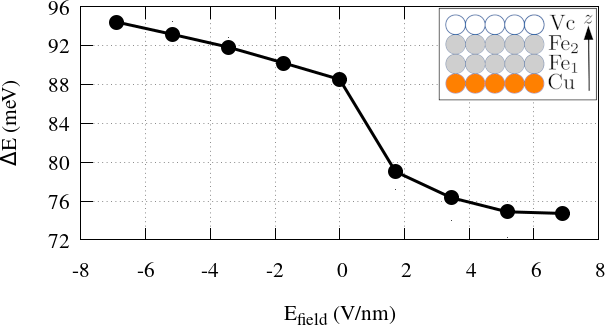}
\caption{\label{fig:gap-bilay-Cu001}Energy gap between the high- and
  low-energy magnon branches at $\bm{q} = \overline{\Gamma}$ for an iron
  bilayer on Cu(001)
  (cf.~Fig.~\protect\ref{fig:CuFe-bilay-eigenmodes}) evaluated as a
  function of E$_{\textrm{field}}$.  }
\end{figure}

\begin{figure}[htb]
 \centering
\includegraphics[width=8.5cm]{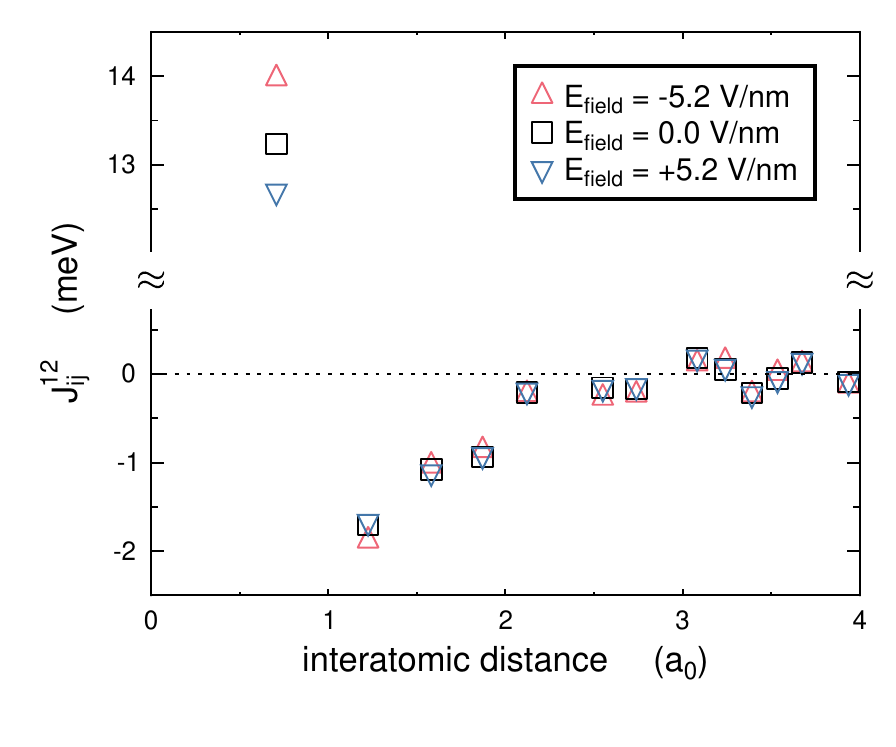}
\caption{\label{fig:CuFe-bilay-J12-QCHRLAY}Inter-layer Heisenberg exchange
  couplings $J^{12}_{IJ}$ for a Fe bilayer on Cu(001) plotted as a function
  of the $|\bm R_I - \bm R_J|$ distance, for 
  E$_{\textrm{field}}$= -5.2, 0, and +5.2
  (V/nm). 
  }
\end{figure}

We then evaluate again the gap
$\Delta E = \omega_2(\overline{\Gamma})-\omega_1(\overline{\Gamma})$
between the high- and low-energy magnon branches 
as a function of E$_{\textrm{field}}$.
For the suspended bilayer its influence was symmetric with respect to the polarity and quite small (Fig.~\ref{fig:gap-bilay-suspended}).  
The presence of the substrate changes the situation dramatically, 
as it can be seen in Fig.~\ref{fig:gap-bilay-Cu001}: the
total variation of $\Delta E$ is now about 30~\% (in contrast with
0.05~\% for the case of a bilayer suspended in vacuum, see
Sec.~\ref{sec:suspend-mono-bilay}) and it is asymmetric with
respect to \eff.  
This outcome is not only due to the different effect
of the perturbation on the magnetic moments for Fe$_1$ and Fe$_2$ atoms (see Fig.~\ref{fig:SMT-QCHRLAY-bilay-Cu001}) but it is also due to the E$_{\textrm{field}}$-induced modifications of the
interlayer Heisenberg exchange couplings \cite{Mankovsky2021a}.  This
can be seen in Fig.~\ref{fig:CuFe-bilay-J12-QCHRLAY} where we present
the inter-layer coupling constants $J^{12}_{IJ}$, for different values
of the external electric field.  The largest variation occurs among the nearest-neighbors and then decays rapidly with the distance $|\bm R_I - \bm R_J|$.


\section{\label{sec:discuss}Discussion}

The calculations presented in Sec.~\ref{sec:numerical-results} reveal 
that certain features of magnon spectra can be controlled 
by an applied electric field, beside aspects already considered in the
literature as a consequence of voltage-controlled magneto-crystalline anisotropy \cite{Rado1979,Liu2013},
multiferroic coupling \cite{Rovillain2010,Risinggard2016},
induced effective DMI \cite{Liu2011b,Zhang2014a,Krivoruchko2017a,Krivoruchko2018,Rana2019,Savchenko2019b,Krivoruchko2020},
or strain from a piezoelectric substrate \cite{Qin2021}.
In particular, we see that a finite E$_{\textrm{field}}$ perturbation
may lead to sizable changes in the magnon lifetime,
even in a case for which the adiabatic dispersion $\omega(\bm{q})$ is fairly unaffected 
(compare Fig.~\ref{fig:magnons-QCHRLAY-mono-Cu001} with Fig.~\ref{fig:damping-QCHRLAY-38-CuFe}).
The stability of this latter quantity can be linked 
to the balance between the tiny asymmetric increase 
of the spin magnetic moment for $|E_{\textrm{field}}| > 0$ on the one hand
(Fig.~\ref{fig:SMT-QCHRLAY-CuFe001}),
and the strengthening of Heisenberg $J_{ij}$ parameters (by few tenths of meV)
for nearest-neighbor Fe atoms on the other hand. 

The robustness of $\omega(\bm{q})$ against \eff\ suggests that
the main reason why the magnon lifetime changes with \eff\ is that 
the Bloch spectral functions entering Eq.~\eqref{eq:Stoner-convolution} are
significantly modified by the electric field.
A negative E$_{\textrm{field}}$ couples mainly with minority electronic states, just below the Fermi level (Fig.~\ref{fig:BSF-Fe-mono-Cu001} top).  
This results in more minority states appearing closer to the Fermi
level, with a shift of the $n^{\downarrow}_{\textrm{Fe}}(E)$ bump
toward higher energy from its original position at around $E=-250$ meV
(Fig.~\ref{fig:DOS-QCHRLAY-CuFe001}). 
The net result is an increase in Stoner intensity, which is shown in Fig.~\ref{fig:damping-QCHRLAY-38-CuFe} (bottom) as a noteworthy enhancement of Landau damping at every depicted $\bm q$-point.
An opposite shift of the electronic spectral weight, i.e., to lower
energies, takes place for $E_{\textrm{field}} > 0$. This results in
longer magnon lifetimes due to the repulsion to deeper energies of the
same minority electronic states discussed above, until they are pushed
below the $[E_{\textrm{min}},E_{\textrm{max}}]$ energy interval
sampled by Eq.~\ref{eq:Stoner-convolution},
and progressively allow only fewer competing Stoner excitations.

For both electric field polarities, saturation of the change in Landau damping appears when the perturbation 
no longer can redistribute spin-polarized spectral weight
within the energy interval spanned by the magnon.

The scenario of a Fe bilayer on Cu(001) shows E$_{\textrm{field}}$-induced changes in the magnon dispersion relations even before considering finite lifetime effects.
Interestingly, the dependence of the magnetic moments on \eff\ exhibits different trends for each of the two iron sheets (see Fig.~\ref{fig:SMT-QCHRLAY-bilay-Cu001}).  In both cases, the magnetic moment is larger than in bulk bcc
Fe, as it is common for surfaces. 
This is a consequence of the thin film straining to follow the different lattice parameters of the substrate. In addition, the reduced dimensionality, 
or more specifically, the reduced number 
of Fe atoms with alike neighbours also plays a role.
%
%
However, whereas the surface Fe$_2$
layer shows an approximately parabolic and slightly
asymmetric variation of the spin magnetic moment with \eff, 
similar to the case of a monolayer  (cf.~Fig.~\ref{fig:SMT-QCHRLAY-CuFe001}), 
the sub-surface Fe$_1$ layer contiguous to copper shows a monotonous
quasilinear dependence instead.  
It seems that exposition to the electric field perturbation 
with or without an in-between layer that can provide metallic screening
is more important than the proximity to the non-magnetic substrate,
in governing these trends.

After the non-magnetic Cu(001) substrate has lifted the degeneracy between the two iron sheets, our calculations show in Fig.~\ref{fig:SMT-QCHRLAY-bilay-Cu001} different trends for the magnetic moment dependence on E$_{\textrm{field}}$
from sub-surface Fe$_1$ contiguous to copper,
and from exposed Fe$_2$ facing vacuum.
The change spans an alike interval of about $0.012$ $\mu_B$.
%
The deeper iron sheet shows an approximately parabolic and slightly asymmetric variation in the spin magnetic moment, similar to the monolayer case of Fig.~\ref{fig:SMT-QCHRLAY-CuFe001}.
The variation is linear instead for the surface Fe$_2$ atoms.

For all cases under consideration we find a $\omega_1(\bm q)$ solution to Eq.~\eqref{eq:Landau-Lifschitz-secular} 
that requires zero energy at the $\overline{\Gamma}$ point, 
i.e. a Goldstone mode.
The second eigenmode $\omega_2(\bm q)$, when present, 
starts from the origin of the Brillouin zone in similar quadratic fashion,
which is a consequence of the ferromagnetic ground state order.
While small-wavelength magnons are equally hosted by both layers,
in the presence of a copper substrate
the two modes 
are neither degenerate in energy, 
nor in the way that they involve Fe atoms from one or the other sheet
at large $\bm q$.

Upon including a finite electric field,
the Goldstone theorem continues to apply and
the lower-energy $|\omega_1(\bm q)\rangle$ branch
continues to start from zero energy. 
The $\Delta E$ gap at $\overline{\Gamma}$ 
strongly depends on the presence of the non-magnetic substrate (cf. Fig.~\ref{fig:gap-bilay-suspended} vs. Fig.~\ref{fig:gap-bilay-Cu001}).
In this case the applied perturbation
significantly
modifies the higher-energy $\omega_2(\bm q = \overline{\Gamma})$ solution,
by changing both 
the inter-layer Heisenberg exchange parameters $J_{IJ}^{12}$,
and layer-resolved magnetic moment $m_1^z$, $m_2^z$
that enter Eq.~\eqref{eq:Landau-Lifschitz-bilayer}.
The resulting energy difference 
gets wider for negative E$_{\textrm{field}}$,
and shrinks but remains open 
when inverting the sign of the perturbation.
A negative electric field 
not only increases the spin magnetic moment 
of both Fe$_1$ and Fe$_2$ atoms which are equally involved 
in the $\omega_n(\bm q \to \overline{\Gamma})$ limit, 
but it also strengthens the $J_{ij}^{12}$ inter-layer interaction (Fig.~\ref{fig:CuFe-bilay-J12-QCHRLAY}).
The opposite happens for $E_{\textrm{field}} > 0$. 

In summary, the electric field perturbation
acts across the dielectric barrier of Fig.~\ref{fig:device-layout}
by modulating the influence of the non-magnetic substrate.
This mechanism provides different Landau damping 
even for limited changes in the purely adiabatic dispersion relation 
of magnons in simple metallic thin films.
The same mechanism also offers possible routes 
to engineer specific changes in the magnon spectrum of more complex, thicker 2D systems, 
such as the energy gap 
at the $\overline{\Gamma}$ point.

We have focused here on simple examples with a ferromagnetic ground state.
However, analogous considerations should apply to more complex
scenarios, such as antiferromagnets \cite{Cheng2016,Wang2020a,Kim2018a}, skyrmion lattices \cite{Chen2019c}, rare earths \cite{Leon2017}, or cases where the
applied electric field is spatially inhomogeneous
\cite{Krivoruchko2019,Krivoruchko2019a}.

%


\section{\label{sec:conclusions}Conclusions}

Magnon spectra of magnetic/non-magnetic metallic heterostructures can
be manipulated by external gating electric field.  Our ab-initio
calculations for test systems of a Fe monolayer and a Fe bilayer, both
suspended in vacuum and deposited on Cu(001), demonstrate that
this perturbation can induce sizable modifications 
in finite magnon lifetimes from Landau damping, 
beside possible changes in the purely adiabatic dispersion
relations already considered in the literature.
The changes in magnon lifetimes can be related to modifications 
of the electronic structure, in particular in the layer-resolved
spin-polarized Bloch spectral functions.  

For systems with more magnon dispersion branches, 
variation of the gap between high- and low-energy
eigenmodes with the external field \eff\ can be expected.
As the E$_{\textrm{field}}$ perturbation controls the degree of
hybridization among magnetic/non-magnetic layers, one can expect
considerable variability in how the magnon spectra are affected by the
external field, depending on the choice of the substrate and the
thickness of the magnetic film.

\section{Acknowledgments}
We gratefully acknowledge computational resources from the Information Technology for Innovation (IT4I) grants: OPEN-19-45 and OPEN-22-40 (Czech National Computing Centre, Ostrava, Czech Republic).
Part of this work was supported by the Deutsche Forschungsgemeinschaft via the grant: DFG EB 154/35,
by the Czech Science Foundation via the grant EXPRO no. 19-28375X, 
and by the Czech Minisitry of Education, Youth and Sports via the grant:
CEDAMNF CZ.02.1.01/0.0/0.0/15\_003/0000358 (Computational and Experimental Design of Advanced Materials with New Functionalities).

\bibliography{liter-magnons}

\end{document}